\documentclass[pdflatex,sn-mathphys]{sn-jnl}

\jyear{2022}%

\theoremstyle{thmstyleone}%
\theoremstyle{thmstyletwo}%

\theoremstyle{thmstylethree}%

\usepackage{lipsum}
\usepackage{amsfonts}
\usepackage{bbm}
\usepackage{textcomp}
\usepackage{bm}
\usepackage{mathrsfs}
\usepackage{multirow}
\usepackage{xcolor}
\usepackage{mathtools}
\usepackage[utf8]{inputenc}

\newcommand{\varhyphen}[1]{{\operatorname{\mathit{#1}}}}

\let\proglang=\textsf

\begin{document}

\title[The hypergeometric test performs comparably to TF-IDF]{The hypergeometric test performs comparably to TF-IDF on standard text analysis tasks}


\author*[1]{\fnm{Paul} \sur{Sheridan}}\email{paul.sheridan.stats@gmail.com}

\author[2]{\fnm{Mikael} \sur{ Onsj{\"o}}}

\affil*[1]{\orgdiv{School of Mathematical and Computational Sciences}, \orgname{University of Prince Edward Island}, \orgaddress{\street{550 University Ave}, \city{Charlottetown}, \postcode{C1A 4P3}, \state{Prince Edward Island}, \country{Canada}}}

\affil[2]{Independent Researcher}

\abstract{Term frequency–inverse document frequency, or TF–IDF for short, and its many variants form a class of term weighting functions the members of which are widely used in text analysis applications. While TF–IDF was originally proposed as a heuristic, theoretical justifications grounded in information theory, probability, and the divergence from randomness paradigm have been advanced. In this work, we present an empirical study showing that TF–IDF corresponds very nearly with the hypergeometric test of statistical significance on selected real-data document retrieval, summarization, and classification tasks. These findings suggest that  a fundamental mathematical connection between TF–IDF and the negative logarithm of the hypergeometric test P-value (i.e., a hypergeometric distribution tail probability) remains to be elucidated. We advance the empirical analyses herein as a first step toward explaining the long-standing effectiveness of TF–IDF from a statistical significance testing lens. It is our aspiration that these results will open the door to the systematic evaluation of significance testing derived term weighting functions in text analysis applications.}




\maketitle

\section{Introduction}\label{sec:intro}

There are two broad questions that text analysis practitioners are wont to ask about a collection of documents. The first is: What documents are most relevant to a given query consisting of one or more terms? This is the problem of document retrieval. The second is: What terms in a given document best characterize its subject matter? This is the problem of document summarization, which is a springboard into document classification and document clustering.

We consider a bag-of-words model setting. Let $\mathbb{D} = \left\{ d_1, d_2, \ldots, d_{N}\right\}$ be a set of~$N$ documents such that each $d_i \in \mathbb{D}$ is a multiset of terms taken from the $M$ term vocabulary $\mathbb{T} = \left\{ t_1, t_2, \ldots, t_M \right\}$. The specification of an $M \times N$ \emph{term-document matrix} $\bm{S} = [s_{ij}]$ for a given set of documents is the textbook starting point for any analysis. The entries of $\bm{S}$ define a \emph{scoring function}, $s:\mathbb{T} \times \mathbb{D} \rightarrow \mathbb{R}_{\ge 0}$. Each entry $s_{ij}$ ($1 \leq i \leq M$ and $1 \leq j \leq N$) is a nonnegative score reflecting the importance of term~$t_i$ to document~$d_j$. The more highly a term scores in a document, the more representative it is considered to be of the document's content.

This conceptualization of a set of documents allows us to recast our opening pair of questions in technical terms. The document retrieval problem amounts to using term-document matrix scores to calculate a ranking for the documents in $\mathbb{D}$ (corresponding to the columns of $\bm{S}$) according to their relevance to a user submitted query $q = \left\{ t_{l_1}, t_{l_2}, \ldots, t_{l_m} \right\}$ of $1 \leq m \leq M$ terms from~$\mathbb{T}$ (one or more rows of $\bm{S}$). By contrast, document summarization or classification can be achieved by using the top $1 \leq m \leq M$ scoring terms in~$\mathbb{T}$ (one or more rows of $\bm{S}$) as a proxy for the subject matter of a given document of interest from~$\mathbb{D}$ (a column of $\bm{S}$).

\subsection{Term frequency–inverse document frequency}

Prerequisite to analysis is the adoption of a concretely defined scoring function. Introduced by Salton and Yang~\cite{Salton1973}, the \emph{term frequency–inverse document frequency} (TF–IDF) scoring function has been a mainstay of information retrieval science for nearly half a century~\cite{Rathi2023}. It scores term~$t_i \in \mathbb{T}$ in document $d_j \in \mathbb{D}$ in accordance with the formula
\begin{equation} \label{eq:tfidf}
\varhyphen{TF-IDF}(t_i,d_j) = k_{ij} \times \log(N/K_{i})\,
\end{equation}
where the \emph{term frequency} (TF) $k_{ij}$ is the number of times $t_i$ occurs in $d_j$, and~$K_i$ the number of documents in~$\mathbb{D}$ containing at least one occurrence of the term $t_i$. The name \emph{inverse document frequency} (IDF) is given to the factor $\log(N/K_i)$. The intuition behind TF–IDF is that a term which is disproportionately concentrated in a few documents tends to score more highly in those documents than do terms occurring frequently in many documents such as articles, prepositions, and conjunctions. Hence a document's highest TF–IDF scoring terms can ordinarily be expected to serve as an informative characterization of its subject matter.

\subsection{Motivation}

Although TF–IDF has traditionally been framed as a heuristic~\cite{Robertson2004}, a series of efforts have been undertaken to explain its empirical success within one or another theoretical framework~\cite{Hiemstra2000, Amati2002, Aizawa2003, deVries2005, Elkan2005, Roelleke2006, Roelleke2008, Roelleke2013, Havrlant2017}. In this paper we explore using the framework of statistical significance testing to gain novel insights into why TF–IDF works. In particular, we show that TF–IDF performs comparably to the hypergeometric significance test for over-representation, hereafter referred to as the hypergeometric test. The test is identical to the one-tailed version of Fisher's exact test~\cite{Rivals2007}. By presenting empirical evidence of a marked agreement between TF–IDF and the hypergeometric test, we hope to provide the multimedia analysis practitioner not only with a deeper understanding of TF–IDF, but also promote with them the notion that the application of statistical significance testing based term weighting functions in cross-modal analyses remains to be explored.

\subsection{The hypergeometric test}

The hypergeometric test is commonly used in bioinformatics research to identify statistically over-represented genes in lists of genetic pathways~\cite{Boyle2004, Huang2009, Maere2005, WardeFarley2010, Zheng2008}. It can be applied in our present setting to assess the claim that the proportion $p_1$ that term $t_i$ makes up of the terms of $d_j$ (cf.~$\hat{p}_1 = k_{ij}/n_j$) is at least as great as the proportion $p_2$ that term $t_i$ makes up of the terms of $\mathbb{D}\setminus d_j$ (cf.~$\hat{p}_2 = (\mathcal{K}_i - k_{ij})/(\mathcal{N} - n_j$)), where term $t_i$ occurs $\mathcal{K}_i = \sum_{j=1}^N k_{ij}$ times in all and $\mathcal{N} = \sum_{j=1}^N n_j$ is the total number of terms. The P-value associated with observing an outcome at least as extreme as the null hypothesis $\mathcal{H}_0: p_1 = p_2$ is given by the hypergeometric distribution tail probability
\begin{equation} \label{eq:hgt}
P(k_{ij};n_j,\mathcal{K}_i,\mathcal{N}) = \sum_{k = k_{ij}}^{n_j} \frac{\binom{\mathcal{K}_i}{k} \binom{\mathcal{N}-\mathcal{K}_i}{n_j-k}}{\binom{\mathcal{N}}{n_j}}\,.
\end{equation}
Framed in the language of the hypergeometric test, Eq.~(\ref{eq:hgt}) represents the chance of drawing term $t_i$ at least $k_{ij}$ times out of $n_j$ draws from a population of $\mathcal{N}$ terms out of which term~$t_i$ occurs $\mathcal{K}_i$ times. The term $t_i$ is said to be \textit{over-represented} in document $d_j$ with respect to the background set of documents $\mathbb{D}$, if the calculated P-value falls short of a preselected significance level. Defining the related hypergeometric test scoring function as the negative logarithm of Eq.~(\ref{eq:hgt}), that is,
\begin{equation} \label{eq:hgt_score}
HGT(t_i, d_j) = -\log\left(\sum_{k = k_{ij}}^{n_j} \frac{\binom{\mathcal{K}_i}{k} \binom{\mathcal{N}-\mathcal{K}_i}{n_j-k}}{\binom{\mathcal{N}}{n_j}}\right) \,,
\end{equation}
ensures that it takes on the same range of values as does TF–IDF; namely, values in the interval $[0,\infty)$. Note that Eq.~(\ref{eq:hgt_score}) preserves term rankings since the negative logarithm transformation is monotonic.

\subsection{Key contributions}

There are multiple ways to explain why TF–IDF works well in practice. Our primary contribution in this paper is to advance empirical evidence suggesting a novel theoretical underpinning for TF–IDF based in statistical significance testing awaits formal elucidation. We also demonstrate that the hypergeometric test can outperform TF–IDF at some typical document retrieval and document classification tasks, hinting that exploring the use of more elaborate tests of statistical significance in text analysis and multimedia studies to be a worthwhile pursuit.

\subsection{Organization of the paper}

Following a brief review of TF–IDF and its existing theoretical justifications in Section~\ref{sec:related_work}, we take to comparing the performance of the TF–IDF scoring function with that of the hypergeometric test based one in a real data setting in Section~\ref{sec:case_studies}. Subsections~\ref{subsec:data_preprocessing} and~\ref{subsec:experimental_setup} detail any data preprocessing steps to which we subjected the data and our general experimental setup, respectively. In Subsection~\ref{subsec:nysk_case_study}, we carried out document retrieval and document summarization experiments on a real-world text collection consisting of roughly $10,000$ online English news articles pertaining to the highly publicized criminal case New York~v.~Strauss-Kahn~\cite{Dermouche2014}. We present in Subsection~\ref{subsec:cranfield_case_study} an ad-hoc document retrieval task analysis on the classical Cranfield 1400 test collection~\cite{GlasgowIRG2023}. The final of our three case studies, found in Subsection~\ref{subsec:20_newsgroups_case_study}, is a document classification task analysis performed on the 20~Newsgroups dataset~\cite{Lang1995}. In each case, we found that TF–IDF consistently produces results very similar to those obtained with the hypergeometric test, and that the degree of agreement between the two scoring functions cannot be accounted for by the effect of the TF term alone. Section~\ref{sec:discussion} concludes the paper with a recapitulation of our main findings and a discussion of some ideas for future work.

\section{Related work}\label{sec:related_work}

\subsection{TF–IDF in text analysis and beyond}

Term weighting functions based on TF–IDF and its variants have long been a fixture of text analysis algorithms~\cite{Rathi2023}. TF–IDF-like weighting function have been notably developed and deployed in the context of document retrieval~\cite{Robertson1994, Robertson2009, Lv2011, Jiminez2018}, document classification~\cite{Joachims1997, Sabbah2017, Kim2019, Jiang2021, Chawla2023}, document clustering~\cite{Bafna2016, Marcinczuk2021, Thielmann2023}, and document summarization in the form of keyword extraction~\cite{Firoozeh2020, Koloski2021, Qian2021}.

Although originally developed for text analysis, term weighting functions have factored prominently in image analysis applications. Notably, features extracted from digital images with TF-IDF-based weightings have been exploited in image indexing~\cite{Magdy2020}, image classification~\cite{Yang2007, Moulin2010}, image retrieval~\cite{Suzuki2008, Kondylidis2018}, and duplicate image detection~\cite{Chum2008}. A recent review of image forgery detection techniques by Kaur~\emph{et~al.}~\cite{Kaur2022} highlights feature extraction as a key step in statistical-based forgery detection methods, leaving open the possibility of using TF-IDF-based extracted features in future methods or the enhancing state-of-the-art methods~\cite{Shan2019, Koul2022, Walia2022} with such features. The same can be said for the state-of-the-art in image analysis generally~\cite{Bansal2021a, Bansal2021b, Shaheed2022}.

There is precedent for exploiting TF–IDF-based term weighting functions in multimedia computing applications. Notably, cross-modal text-image retrieval methods incorporating TF–IDF-based weightings have been proposed~\cite{Arensia2012, Schneider2022, Xie2022}. Automatic image captioning is another cross-modal (i.e., text and image) application for which TF–IDF-derived features have been successfully employed~\cite{Pavlopoulos2019, Ashwath2021}. Other prominent cross-modal contexts in which TF–IDF and its variants have been explored include fake news detection~\cite{Masciari2020, Choras2021}, film genre classification~\cite{Mangolin2022, Rajput2022}, and temporal video scene segmentation~\cite{Giveki2021, Kannao2022}.

Transformer-based language models, as exemplified by BERT~\cite{Devlin2019}, constitutes the state-of-the-art when it comes to generating features from text~\cite{vonderMosel2022}. The development of methods that combine classical TF–IDF with transformer-based language model features is an active area of research~\cite{Rathi2023}.

\subsection{TF–IDF theoretical foundations}

There is an appreciable body of scholarship on TF–IDF theoretical foundations. Robertson \cite{Robertson2004} reviews some of the more notable attempts to place the IDF metric on a sound theoretical footing. His paper is, moreover, a fair starting-point for learning the layout of the theoretical foundations for TF–IDF landscape. However it is Thomas Roelleke who has emerged as the primary authority on the subject. Roelleke in essence identifies four types of theoretical argument for TF–IDF~\cite{Roelleke2013}: those based on information theory, those based on probabilistic relevance modelling, those based on statistical language modelling, and those based on divergence from randomness models.

\subsubsection{Information theoretic approaches}

Aizawa \cite{Aizawa2003} hit on a connection between mutual information and TF–IDF by building on the earlier work~\cite{Robertson1974, Wong1992} to place IDF in an information theoretic framework. Robertson~\cite{Robertson2004} writes that ``it is difficult to see it [Aizawa's analysis] as an explanation of the value of IDF''~(p.~508), leading one to presume his difficulty carries over to TF–IDF by extension. His criticism almost certainly inspired Roelleke to seek out a logically incontestable explanation for TF–IDF within the realm of information theory. We are happy to report that Roelleke was wholly successfully in this effort. For in~\cite{Roelleke2013}, he convincingly relates a form of TF–IDF to the difference between two Kullback--Leibler divergences.

\subsubsection{Probabilistic relevance modelling approaches}

A number of historically influential document retrieval models and their many variants were devised within the framework of probabilistic relevance modelling~\cite{Robertson1976}. heoretical justifications for IDF have been arrived at starting from one or another elaboration on the Binary Independence Model of probabilistic relevance~\cite{Robertson2004, deVries2005, Manning2008}. Of these, only de Vries and Roelleke~\cite{deVries2005} go on to draw a connection, albeit a tenuous one, between TF–IDF and a common Binary Independence Model variant. The Poisson Model is a straightforward count data modelling Binary Independence Model generalization~\cite{Roelleke2013}. Roelleke~\cite{Roelleke2006} has shown under which conditions TF–IDF emerges from the Poisson Model. This marked the first convincing derivation of TF–IDF within anything resembling a classical probabilistic framework. The classic Okapi BM25 scoring function is a generalized version of TF–IDF~\cite{Robertson2004}. The connection between TF–IDF and Okapi BM25 has been elucidated by Roelleke~\cite{Roelleke2008, Roelleke2013}. Wu~\emph{et~al.}~\cite{Wu2008} interpret TF–IDF as a special case of a probabilistic relevance model of their own contrivance. Finally, Joachims~\cite{Joachims1997} presents a rough analysis of TF–IDF within the context of probabilistic relevance modelling-based document classification.

\subsubsection{Statistical language modelling approaches}

Hiemstra~\cite{Hiemstra2000} advanced the first statistical language modelling-based interpretation of TF–IDF. Roelleke and Wang~\cite{Roelleke2008} improve upon Hiemstra's finding by identifying the conditions under which a decidedly more TF–IDF-like formula emerge from statistical modeling language assumptions. In later work, Roelleke~\cite{Roelleke2013} derives TF–IDF as a limiting case of a generalization of a model proposed by Hiemstra. Elkan~\cite{Elkan2005} teases out a loose connection between TF–IDF and the Dirichlet-multinomial distribution Fisher kernel for modeling term burstiness. Roelleke~\cite{Roelleke2013} subsequently established a more direct relationship between TF–IDF and the statistical language modelling of burstiness.

\subsubsection{Divergence from randomness approaches}

Havrlant and Kreinovich~\cite{Havrlant2017} derive TF–IDF (at least approximately) from a simple probabilistic model with statistical language model characteristics. What they evidently did not realize is that Amati and Van Rijsbergen~\cite{Amati2002} had already derived virtually the same result in a divergence from randomness modelling context. Roelleke~\cite{Roelleke2013}, in characteristic fashion, sheds further light on the relationship between TF–IDF and divergence from randomness modelling.

\section{Real-world data case studies}\label{sec:case_studies}

We assessed the correspondence between TF–IDF and the hypergeometric test scoring function on real-world dataset for selected document retrieval, summarization, and classification tasks. We availed ourselves of three datasets:
\begin{description}
    \item[The New York v. Strauss-Kahn (NYSK) dataset:] A collection of $10,421$ online English language news articles about a highly publicized criminal case relating to allegations that former International Monetary Fund director Dominique Strauss-Kahn had sexually assaulted a hotel maid. The dataset, as curated by Dermouche \emph{et al.}~\cite{Dermouche2014}, has been made publicly available for download at the UCI Machine Learning Repository website~\cite{Dua2023}.
    \item[The Cranfield 1400 test collection:] A foundational benchmark dataset used in information retrieval research. It consists of $1,398$ aerodynamics article abstracts, $225$ queries, and relevance judgements for all query/document pairs. We analyzed a TREC XML formatted version of the data~\cite{Oussama2022}. The original dataset is publicly available and can be downloaded at the Glasgow Information Retrieval Group website~\cite{GlasgowIRG2023}.
    \item[The 20 Newsgroups dataset:] A collection comprising $20,000$ Usenet newsgroup posts on $20$ topics that is commonly used for comparing different text classification methods. The collection was curated by Ken Lang~\cite{Lang1995}. We analyzed a preprocessed version of the dataset, consisting of a subset of some $18,846$ posts, that is included in the Python library \textbf{sklearn}~1.2.2~\cite{Pedregosa2011}.
\end{description}
In our comparisons below, we consistently find that TF–IDF and the hypergeometric test scoring function agree with each other significantly more than either scoring function agrees with a TF scoring function baseline. Comparing hypergeometric test performance to state-of-the-art scoring functions is beyond the scope of this work. Rather, it is our aim to demonstrate that the hypergeometric test yields comparable results to those of the classical TF–IDF scoring function.

\subsection{Data preprocessing}\label{subsec:data_preprocessing}

We subjected the raw NYSK data to a series of routine preprocessing steps as a prelude to analysis. The procedure we implemented in Python~3.10.11. For each article in the collection, we first excised and tokenized the body text, then removed any non-ASCII terms, then converted all terms to lowercase, then removed any punctuation marks, then substituted textual representations for any terms representing integers, then removed all stop words, and finally lemmatized any remaining terms. In addition to standard Python libraries, we made use of the natural language processing libraries \textbf{nltk}~3.8.1, \textbf{BeautifulSoup4}~4.11.2, \textbf{inflect}~6.0.4 and \textbf{contractions}~0.1.73. The total number of unique terms after preprocessing is $74,004$. We applied the same procedure to preprocess the documents and queries of the Cranfield collection. The 20~Newsgroups dataset we analyzed directly without the application of any preprocessing steps.

\subsection{Experimental setup}\label{subsec:experimental_setup}

To compare TF–IDF with the hypergeometric test scoring function, we calculated a trio of term-document matrices for each document collection analyzed in this study: one based on the TF–IDF scoring function of Eq.~(\ref{eq:tfidf}), one based on the hypergeometric test scoring function of Eq.~(\ref{eq:hgt_score}), and another based on the TF scoring function for use as a baseline measure. We evaluated the agreement between the TF–IDF and the hypergeometric test scoring function in three different scenarios: document retrieval, summarization, and classification. The questions we provide quantitative answers to are as~follows:
\begin{description}
    \item[Document retrieval scenario:] To what extent do the the TF–IDF and the hypergeometric test scoring functions agree on which documents are most relevant to a given user submitted~query?
    \item[Document summarization scenario:] To what extent do the the TF–IDF and the hypergeometric test scoring functions agree on which among a given document's terms best characterize its subject matter?
    \item[Document classification scenario:] To what extent do the the TF–IDF and the hypergeometric test scoring functions agree on assigning predefined categories to text documents based on their contents?
\end{description}

The NYSK document collection lacks accompanying ground truth data on which to objectively evaluate the performances of our three scoring functions of interest (i.e., TF–IDF, the hypergeometric test scoring function, and the TF baseline). It is nonetheless useful for illustrating how the hypergeometric test scoring function produces similar rankings to those of TF–IDF as regards term-document matrix rows (pertinent to document retrieval) and columns (pertinent to document summarization). In this case study, we took TF–IDF generated rankings as ground truth, and used the \textit{Precision at 10} (P@10) evaluation measure to quantify how well the hypergeometric test scoring function outcomes agree on particular document retrieval and summarization tasks.

The Cranfield collection case study constitutes a standard ad-hoc document retrieval task. In this context, we took each of TF, TF–IDF, and the hypergeometric test scoring function as a retrieval function. For each retrieval function, we ranked collection documents for each query according to the cosine similarity measure. Using associated relevance judgements, we compared the results based on common performance evaluation measures (e.g., P@10, F1 score, and MAP among others). Lastly, we report the percent improvements that come with using the hypergeometric test retrieval function over that of TF–IDF.

The goal of the 20~Newsgroups case study is to show that TF–IDF and the hypergeometric test scoring function yield comparable document classifications. We used the default \textbf{sklearn} split of the documents into training (60\%) and test (40\%) document sub-collections. We trained a multinomial Naive Bayes classifier on features extracted using each of the TF, TF–IDF, and the hypergeometric test scoring functions. We report precision, recall, and F1 scores for TF–IDF and the hypergeometric test as broken down by category, as well as in aggregate. The corresponding TF performance evaluation values we report for the aggregate case alone.

\subsection{NYSK dataset case study}\label{subsec:nysk_case_study}

\subsubsection{Document retrieval scenario}\label{subsubsec:nysk_doc_retrieval}

\begin{figure}[ht]
\begin{center}
\includegraphics[width=119mm]{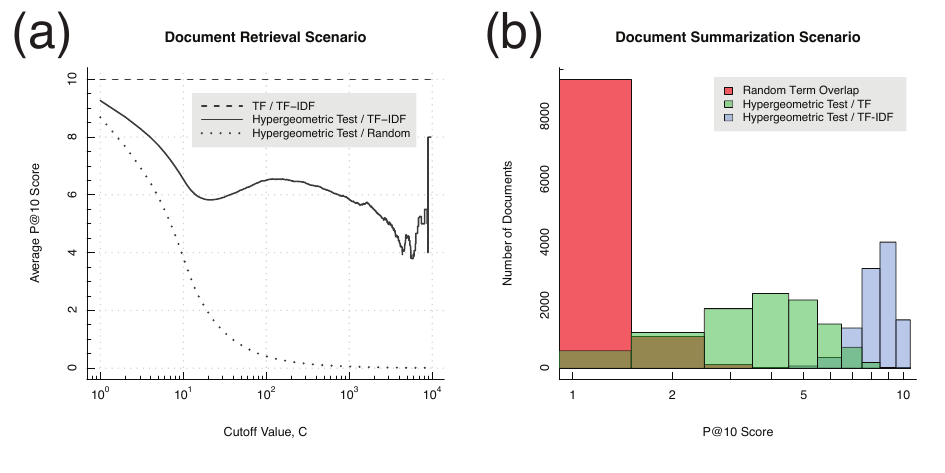}
\end{center}
\caption{The hypergeometric test produces outcomes similar to those of TF–IDF on the NYSK dataset in (a) the document retrieval scenario for one-term queries and (b) the document summarization scenario}\label{fig:nysk_plots}
\end{figure}

We ran one- and two-term queries on the NYSK documents, of which one will recall there are $N= 10,421$ in number, to compare TF–IDF with the hypergeometric test.

We queried the NYSK documents on each of the $M = 74,004$ unique terms making up the dataset. For each one-term query $q =\{t_i \}$ ($1 \leq i \leq M$), we ranked the documents in decreasing order of TF–IDF score, the hypergeometric test scoring function score, and TF score, respectively.

The average P@10 score between {TF–IDF and the hypergeometric test scoring function works out to $6.54\pm 2.36$ in the case when we confine ourselves to those $15,788$ terms occurring in at least ten documents. Contrast with this with the average P@10 score of $3.84\pm 3.10$ that we obtained by calculating P@10 scores between corresponding pairs of randomized document rankings. This shows that the hypergeometric test scoring function and TF–IDF are more similar to each other than either one is to a baseline measure generated from the P@10 scores of random orderings of documents. It is important to note, however, that the average P@10 score between the hypergeometric test scoring function and~TF equals that of the hypergeometric test scoring function and TF–IDF. This is because~TF and TF–IDF give rise to identical document rankings on account that the IDF factor in TF–IDF is constant across all documents for any one-term query. It will come as no surprise, then, when we report an average P@10 score of $10.00\pm 0.00$ between~TF and TF–IDF. 

The plot of Fig.~\ref{fig:nysk_plots}(a) shows an average of one-term query P@10 scores (along the vertical axis) plotted as a function of a cutoff value, $C$, (along the logarithmically scaled horizontal axis). The average P@10 score for a given value of $C$ is calculated by adding up the P@10 scores of precisely those terms occurring in at least $C$ documents and then dividing by the total number of scores. In what preceded we examined the special case when~$C$ is equal to ten. The hypergeometric test/TF–IDF average P@10 scores (solid line) conspicuously exceed those calculated between the hypergeometric test and the random document overlap baseline (dotted line). Note, however, that one-term queries cannot be used to discriminate between~TF and TF–IDF (dashed line). A cursory inspection of the plot reveals that the conclusions we drew in the~$C=10$ case hold generally true, namely:~1) the hypergeometric test scoring function and TF–IDF produce top ten document rankings that are much more similar each other than they are to random orderings of documents, but 2) it is impossible to differentiate TF–IDF from tp on the basis of one-term queries. The latter conclusion is unsatisfying as our objective is to compare the hypergeometric test scoring function with TF–IDF, rather than the na{\"i}ve~TF scoring function.

This brings us to the two-term query case. We are immediately confronted with the problem of how to go about constructing queries that discriminate TF–IDF from TF as much as possible. To meet this challenge we isolated a small pool of ``bursty'' terms from which individual query terms are to be selected. A term is said to be bursty, loosely speaking, when its occurrences are concentrated in very few documents. We evaluated the Irvine and Callison-Burch~\cite{Irvine2017} proposed term burstiness measure
\begin{equation}\label{eq:icb_burstiness}
B(t_i) = \frac{1}{K_i} \sum_{j = 1}^N \frac{k_{ij}}{n_j}
\end{equation}
on those terms occurring in at least ten documents. The top $106$ highest scoring terms we identified as bursty, and separated into two groups based on high/low document proportion, $K_i/N$. The six bursty terms with highest $K_i/N$ value we call \textit{common} (i.e., ``strausskahn,'' ``say,'' ``new,'' ``imf,'' ``comment,'' and ``lagarde''), and the remaining 100 bursty terms we call \textit{rare}. Each common bursty term we paired with each rare bursty terms for a total of six different sets of $100$ two-term queries.

\begin{table}[ht]
\caption{Scoring function comparison results for two-term queries on the NYSK data}\label{tab:two_term_query_results_hgt}
\centering
\begin{tabular}{lccccc}
Experiment 	& $B(t_i)$ 	& Random 		& HGT / TF  		& HGT / TF–IDF 	& TF–IDF / TF \\
\midrule
strausskahn 	& 0.0161 		& $0.01\pm 0.00$ 	& $1.78\pm 2.38$ 	& $8.17\pm 1.94$ 	& $1.95\pm 2.50$ \\
say 			& 0.0161 		& $0.01\pm 0.00$ 	& $2.20\pm 2.72$ 	& $7.87\pm 2.22$ 	& $2.30\pm 2.85$ \\
imf 			& 0.0135 		& $0.01\pm 0.00$ 	& $1.54\pm 2.05$ 	& $7.87\pm 2.23$ 	& $1.55\pm 2.05$ \\
new 			& 0.0095 		& $0.01\pm 0.00$ 	& $1.57\pm 2.21$ 	& $8.08\pm 1.99$ 	& $1.67\pm 2.20$ \\
lagarde 		& 0.0095 		& $0.04\pm 0.00$ 	& $1.22\pm 1.88$ 	& $7.98\pm 2.44$ 	& $1.43\pm 2.10$ \\
comment 		& 0.0091 		& $0.02\pm 0.00$ 	& $1.14\pm 2.04$ 	& $8.41\pm 1.98$ 	& $1.21\pm 2.10$ \\
\midrule
\textbf{Average} 	& 0.0123 		& 0.17 			& 1.57 			& 8.06 			& 1.68
\end{tabular}
\end{table}

Table~\ref{tab:two_term_query_results_hgt} summarizes our findings for the six different two-term query experiments Two-term query scores are evaluated as the sum of the scores of their constituent terms. Each common bursty term doubles as an experiment name (first column). The first column lists in decreasing order of burstiness (second column) the six common bursty terms as determined by the scoring function of Eq.~(\ref{eq:icb_burstiness}). A total of $100$ two-term queries were formed for each experiment by pairing a common bursty term with each of the $100$ rare bursty terms. The rightmost four columns show average P@10 scores between selected scoring functions with standard deviations. The hypergeometric test scoring function agrees with TF–IDF (fifth column) beyond what can be accounted for by the random document overlap baseline (third column). Unlike with the one-term query experiment, the agreement between the hypergeometric test scoring function and TF–IDF is not explained by the TF scoring function alone (fourth and sixth columns).

As with the one-term query experiment, the hypergeometric test scoring function remains comfortably in excess of the agreement observed between the hypergeometric test scoring function and the TF baseline, not to mention the hypergeometric test scoring function and the random document overlap baseline. In the ``strausskahn'' experiment, for instance, we find an average P@10 score of $8.17\pm 1.94$ between TF–IDF and the hypergeometric test scoring function, as compared with $1.95\pm 2.50$ between the hypergeometric test scoring function and the TF baseline. Unlike with the one-term query experiment, the agreement between the hypergeometric test scoring function and TF–IDF is not solely attributable to TF, as the average P@10 score between TF and TF–IDF is a mere $1.78\pm 2.38$. Results from the five other two-term query experiments can be similarly interpreted. Empirical evidence in support of the claim that TF–IDF agrees with the hypergeometric test scoring function beyond what the TF baseline can explain is thus supplied.

\subsubsection{Document summarization scenario}\label{subssubec:nysk_doc_sumarization}

We summarized each NYSK document according to its top ten highest hypergeometric test scoring terms and compared these results with those of our three now familiar alternative scoring functions.

The superimposed histograms of Fig.~1(b) show P@10 scores as measured between the hypergeometric test and each of three alternatives in the document summarization scenario. The blue barred histogram casts into clear relief a notable agreement between the hypergeometric test scoring function and TF–IDF. The average P@10 score is $8.47$ with a standard deviation of $1.04$. It is interesting to observe that this is significantly higher than the average P@10 score of $4.18\pm 1.68$ between the hypergeometric test scoring function and the~TF baseline (green histogram). Lastly, the red barred histogram shows the P@10 score distribution for a simple baseline measure where each document is summarized by ten randomly selected terms. The average P@10 score is $0.58$ with a standard deviation of $0.71$.

To sum up: the hypergeometric test scoring function overwhelmingly agrees with TF–IDF in the document summarization scenario, and this cannot be explained by the effect of the TF baseline function, let alone by lists of randomly selected terms.

\subsection{Cranfield 1400 test collection case study}\label{subsec:cranfield_case_study}

Table~\ref{tab:cranfield_results} compares the performance of the hypergeometric test retrieval function with those of TF–IDF and TF on the Cranfield test collection. Consistent with our NYSK dataset document retrieval case study results, the hypergeometric test performance (fourth column) is comparable to the performance of TF–IDF (third column) across a robust set of common evaluation measures (first column). The TF retrieval function baseline performances (second column) show that the conspicuous level of agreement between the hypergeometric test and TF–IDF is not attributable to the TF factor alone. Interestingly, five of the seven evaluation measures have the hypergeometric test outperforming TF–IDF as measured by percent improvement (fifth column). Take the P@10 measure which has the hypergeometric test outperforming TF–IDF by $8\%$. This is decidedly not, however, good evidence that the hypergeometric test is will generally outperform TF–IDF at ad-hoc document retrieval tasks. The Cranfield collection is ill-suited for evaluating this claim. Rather, we used this collection as a toy dataset to study how the hypergeometric test compares to TF–IDF.

\begin{table}[ht]
\caption{Retrieval function comparison results on the Cranfield collection}\label{tab:cranfield_results}
\centering
\begin{tabular}{lcccc}
Evaluation measure & TF & TF–IDF & HGT & Improvement \\
\midrule
F1 score 			& 0.2199 & 0.2549 & 0.2650 & 3.95\% \\
MAP 			& 0.1472 & 0.1900 & 0.1993 & 4.86\% \\
GMAP 			& 0.0731 & 0.1047 & 0.1098 & 4.94\% \\
Reciprocal rank 	& 0.3256 & 0.3899 & 0.4174 & 7.07\% \\
P@10 			& 0.1200 & 0.1613 & 0.1742 & 7.99\% \\
P@50 			& 0.0571 & 0.0679 & 0.0676 & -0.44\% \\
P@100 			& 0.0384 & 0.0436 & 0.0432 & -0.92\% \\
\midrule
\end{tabular}
\end{table}

\subsection{The 20 Newsgroups dataset case study}\label{subsec:20_newsgroups_case_study}

We now turn to the 20~Newsgroups dataset document classification task case study. Table~\ref{tab:20_newsgroups_results} shows the results of running our multinomial Naive Bayes classifier on the data based on using TF–IDF and hypergeometric test extracted features, respectively. For each of the~20 different document classes (first column), the TF–IDF precision, recall, and F1~scores (second to fourth columns) generally agree with the corresponding hypergeometric test scores (fifth to seventh columns) up to a few percentage points. For instance, the ``comp.graphics'' document class (second row) has the hypergeometric test with percent increases over TF–IDF of $4.84\%$ precision-wise, $0.00\%$ recall-wise, and $1.45\%$ by F1~score. Only the ``comp.os.ms-windows.misc'' document class (third row) is observed to buck this general tendency with corresponding percent increases of $18.92\%$ (precision), $475.00\%$ (recall), and $350.00\%$ (F1~score). The main take away, however, is that the hypergeometric test performs comparably to TF–IDF in the aggregate when the evaluation metrics are averaged over the 20 document classes (bottom row). The corresponding evaluation metrics for the TF baseline are $0.79$ (precision), $0.68$ (recall), and $0.67$ (F1~score). As with the previous case studies the agreement between TF–IDF and the hypergeometric test is not entirely explained by the baseline TF scoring function.

\begin{table}[ht]
\caption{Multinomial Naive Bayes classifier results for the 20 Newsgroups dataset with both TF–IDF and hypergeometric test extracted features}\label{tab:20_newsgroups_results}
\centering
\begin{tabular}{lcccccc}
\multicolumn{1}{c}{} & \multicolumn{3}{c}{\textbf{TF–IDF}} & \multicolumn{3}{c}{\textbf{HGT}} \\
\cmidrule(rl){2-4} \cmidrule(rl){5-7}
\textbf{Document class} & {Precision} & {Recall} & {F1} & {Precision} & {Recall} & {F1}\\
\midrule
alt.atheism 				& 0.81 & 0.82 & 0.82 & 0.80 & 0.81 & 0.81 \\
comp.graphics 				& 0.62 & 0.77 & 0.69 & 0.65 & 0.77 & 0.70 \\
comp.os.ms-windows.misc 	& 0.74 & 0.04 & 0.08 & 0.88 & 0.23 & 0.36 \\
comp.sys.ibm.pc.hardwar 		& 0.53 & 0.78 & 0.63 & 0.56 & 0.78 & 0.65 \\
comp.sys.mac.hardware 		& 0.73 & 0.85 & 0.79 & 0.76 & 0.85 & 0.80 \\
comp.windows.x 			& 0.78 & 0.76 & 0.77 & 0.81 & 0.76 & 0.79 \\
misc.forsale 				& 0.80 & 0.76 & 0.78 & 0.80 & 0.78 & 0.79 \\
rec.autos 					& 0.87 & 0.92 & 0.89 & 0.87 & 0.92 & 0.89 \\
rec.motorcycles 			& 0.93 & 0.96 & 0.94 & 0.93 & 0.96 & 0.94 \\
rec.sport.baseball 			& 0.95 & 0.94 & 0.94 & 0.95 & 0.94 & 0.95 \\
rec.sport.hockey 			& 0.96 & 0.97 & 0.96 & 0.96 & 0.97 & 0.96 \\
sci.crypt 					& 0.87 & 0.92 & 0.89 & 0.87 & 0.93 & 0.90 \\
sci.electronics 				& 0.77 & 0.76 & 0.76 & 0.77 & 0.76 & 0.77 \\
sci.med 					& 0.90 & 0.82 & 0.86 & 0.89 & 0.82 & 0.86 \\
sci.space 					& 0.87 & 0.90 & 0.89 & 0.87 & 0.90 & 0.89 \\
soc.religion.christian 			& 0.86 & 0.94 & 0.90 & 0.84 & 0.95 & 0.89 \\
talk.politics.guns 			& 0.82 & 0.90 & 0.85 & 0.80 & 0.89 & 0.84 \\
talk.politics.mideast 			& 0.96 & 0.91 & 0.94 & 0.97 & 0.92 & 0.94 \\
talk.politics.misc 			& 0.72 & 0.66 & 0.69 & 0.74 & 0.66 & 0.70 \\
talk.politics.misc 			& 0.70 & 0.64 & 0.67 & 0.74 & 0.63 & 0.68 \\
\midrule
\textbf{Average} & $0.81$ & $0.80$ & $0.79$ & $0.82$ & $0.81$ & $0.81$
\end{tabular}
\end{table}

\section{Discussion}\label{sec:discussion}

In this paper we have advanced evidence in the form of a real-data case studies that TF–IDF performs comparably to the hypergeometric test on standard document retrieval, summarization, and summarization tasks. These findings we present as a first step toward establishing a theoretical foundation for TF–IDF within the framework of statistical significance testing. A hypergeometric test justification for TF–IDF would offer a new and intuitive way of understanding why TF–IDF has proved to be so effective in practice.

The statistical significance testing paradigm is an attractive and general setting for developing novel term weighting functions for use in text and cross-modal analysis applications, and offers a fresh perspective firmly rooted in the bioinformatics tradition~\cite{Boyle2004, Maere2005, Rivals2007, Zheng2008, Huang2009, WardeFarley2010, Cao2014, Cao2017}. It is curious fact that the class of TF–IDF numerical measures has only recently been brought to bear on bioinformatics problem gene over-representation analysis~\cite{Fan2021}. The hypergeometric test has, in turn, received surprisingly little attention from the text analysis and multimedia studies. Apart from the present work, the performance of TF–IDF as compared with that of the hypergeometric test scoring function on text analysis tasks is, to our knowledge, found only in Onsj{\"o} and Sheridan~\cite{Onsjo2020}. An interesting line of future work would be to compare the performance of significance test derived scoring functions with the state-of-the-art in text and cross-modal analyses on large data benchmark collections. Another interesting avenue of research lies in combining significance test derived features with transformer-based language models.

\section*{Declarations}

\begin{itemize}
\item Conflict of interest/Competing interests: The authors declare no conflict of interest/competing interests.
\item Availability of data and materials: The NYSK dataset is available for download at the UCI Machine Learning Repository at \url{https://archive.ics.uci.edu/ml/datasets/NYSK}. The TREC XML formatted version of the Cranfield collection was downloaded from the \textbf{cranfield-trec-dataset} GitHub repository (\url{https://github.com/oussbenk/cranfield-trec-dataset}, Commit Id: 1208e6edfb6cb2527b2c44398d3d8fefd3249144) . The 20~Newsgroups dataset used in this study is included in the Python library \textbf{sklearn}~1.2.2~\cite{Pedregosa2011}.
\item Code availability: \proglang{R} and Python code used to analyze the data is available at \url{https://github.com/paul-sheridan/hgt-tfidf}.
\end{itemize}

\bibliography{bibliography}


\begin{thebibliography}{75}
\ifx \bisbn   \undefined \def \bisbn  #1{ISBN #1}\fi
\ifx \binits  \undefined \def \binits#1{#1}\fi
\ifx \bauthor  \undefined \def \bauthor#1{#1}\fi
\ifx \batitle  \undefined \def \batitle#1{#1}\fi
\ifx \bjtitle  \undefined \def \bjtitle#1{#1}\fi
\ifx \bvolume  \undefined \def \bvolume#1{\textbf{#1}}\fi
\ifx \byear  \undefined \def \byear#1{#1}\fi
\ifx \bissue  \undefined \def \bissue#1{#1}\fi
\ifx \bfpage  \undefined \def \bfpage#1{#1}\fi
\ifx \blpage  \undefined \def \blpage #1{#1}\fi
\ifx \burl  \undefined \def \burl#1{\textsf{#1}}\fi
\ifx \doiurl  \undefined \def \doiurl#1{\url{https://doi.org/#1}}\fi
\ifx \betal  \undefined \def \betal{\textit{et al.}}\fi
\ifx \binstitute  \undefined \def \binstitute#1{#1}\fi
\ifx \binstitutionaled  \undefined \def \binstitutionaled#1{#1}\fi
\ifx \bctitle  \undefined \def \bctitle#1{#1}\fi
\ifx \beditor  \undefined \def \beditor#1{#1}\fi
\ifx \bpublisher  \undefined \def \bpublisher#1{#1}\fi
\ifx \bbtitle  \undefined \def \bbtitle#1{#1}\fi
\ifx \bedition  \undefined \def \bedition#1{#1}\fi
\ifx \bseriesno  \undefined \def \bseriesno#1{#1}\fi
\ifx \blocation  \undefined \def \blocation#1{#1}\fi
\ifx \bsertitle  \undefined \def \bsertitle#1{#1}\fi
\ifx \bsnm \undefined \def \bsnm#1{#1}\fi
\ifx \bsuffix \undefined \def \bsuffix#1{#1}\fi
\ifx \bparticle \undefined \def \bparticle#1{#1}\fi
\ifx \barticle \undefined \def \barticle#1{#1}\fi
\bibcommenthead
\ifx \bconfdate \undefined \def \bconfdate #1{#1}\fi
\ifx \botherref \undefined \def \botherref #1{#1}\fi
\ifx \url \undefined \def \url#1{\textsf{#1}}\fi
\ifx \bchapter \undefined \def \bchapter#1{#1}\fi
\ifx \bbook \undefined \def \bbook#1{#1}\fi
\ifx \bcomment \undefined \def \bcomment#1{#1}\fi
\ifx \oauthor \undefined \def \oauthor#1{#1}\fi
\ifx \citeauthoryear \undefined \def \citeauthoryear#1{#1}\fi
\ifx \endbibitem  \undefined \def \endbibitem {}\fi
\ifx \bconflocation  \undefined \def \bconflocation#1{#1}\fi
\ifx \arxivurl  \undefined \def \arxivurl#1{\textsf{#1}}\fi
\csname PreBibitemsHook\endcsname

\bibitem{Salton1973}
\begin{barticle}
\bauthor{\bsnm{Salton}, \binits{G.}},
\bauthor{\bsnm{Yang}, \binits{C.S.}}:
\batitle{On the specification of term values in automatic indexing.}
\bjtitle{Journal of Documentation}
\bvolume{29}(\bissue{4}),
\bfpage{351}--\blpage{372}
(\byear{1973}).
\doiurl{10.1108/eb026562}
\end{barticle}
\endbibitem

\bibitem{Rathi2023}
\begin{barticle}
\bauthor{\bsnm{Rathi}, \binits{R.N.}},
\bauthor{\bsnm{Mustafi}, \binits{A.}}:
\batitle{The importance of term weighting in semantic understanding of text:
  {A} review of techniques}.
\bjtitle{Multimedia Tools and Applications}
\bvolume{82},
\bfpage{9761}--\blpage{9783}
(\byear{2023}).
\doiurl{10.1007/s11042-022-12538-3}
\end{barticle}
\endbibitem

\bibitem{Robertson2004}
\begin{barticle}
\bauthor{\bsnm{Robertson}, \binits{S.}}:
\batitle{Understanding inverse document frequency: On theoretical arguments for
  {IDF}}.
\bjtitle{Journal of Documentation}
\bvolume{60}(\bissue{5}),
\bfpage{503}--\blpage{520}
(\byear{2004}).
\doiurl{10.1108/00220410410560582}
\end{barticle}
\endbibitem

\bibitem{Hiemstra2000}
\begin{barticle}
\bauthor{\bsnm{Hiemstra}, \binits{D.}}:
\batitle{A probabilistic justification for using tf.idf term weighting in
  information retrieval}.
\bjtitle{International Journal on Digital Libraries}
\bvolume{3}(\bissue{2}),
\bfpage{131}--\blpage{139}
(\byear{2000}).
\doiurl{10.1007/s007999900025}
\end{barticle}
\endbibitem

\bibitem{Amati2002}
\begin{barticle}
\bauthor{\bsnm{Amati}, \binits{G.}},
\bauthor{\bsnm{Van~Rijsbergen}, \binits{C.J.}}:
\batitle{Probabilistic models of information retrieval based on measuring the
  divergence from randomness}.
\bjtitle{ACM Transactions on Information Systems}
\bvolume{20}(\bissue{4}),
\bfpage{357}--\blpage{389}
(\byear{2002}).
\doiurl{10.1145/582415.582416}
\end{barticle}
\endbibitem

\bibitem{Aizawa2003}
\begin{barticle}
\bauthor{\bsnm{Aizawa}, \binits{A.}}:
\batitle{An information-theoretic perspective of tf-idf measures}.
\bjtitle{Information Processing and Management}
\bvolume{39}(\bissue{1}),
\bfpage{45}--\blpage{65}
(\byear{2003}).
\doiurl{10.1016/S0306-4573(02)00021-3}
\end{barticle}
\endbibitem

\bibitem{deVries2005}
\begin{bchapter}
\bauthor{\bparticle{de} \bsnm{Vries}, \binits{A.P.}},
\bauthor{\bsnm{Roelleke}, \binits{T.}}:
\bctitle{Relevance information: A loss of entropy but a gain for {IDF}?}
In: \bbtitle{Proceedings of the 28th Annual International ACM SIGIR Conference
  on Research and Development in Information Retrieval}.
\bsertitle{SIGIR ’05},
pp. \bfpage{282}--\blpage{289}.
\bpublisher{Association for Computing Machinery},
\blocation{New York, NY, USA}
(\byear{2005}).
\doiurl{10.1145/1076034.1076084}
\end{bchapter}
\endbibitem

\bibitem{Elkan2005}
\begin{bchapter}
\bauthor{\bsnm{Elkan}, \binits{C.}}:
\bctitle{Deriving {TF-IDF} as a {Fisher} kernel}.
In: \bbtitle{Proceedings of the 12th International Conference on String
  Processing and Information Retrieval}.
\bsertitle{SPIRE'05},
pp. \bfpage{295}--\blpage{300}.
\bpublisher{Springer},
\blocation{Berlin, Heidelberg}
(\byear{2005}).
\doiurl{10.1007/11575832_33}
\end{bchapter}
\endbibitem

\bibitem{Roelleke2006}
\begin{bchapter}
\bauthor{\bsnm{Roelleke}, \binits{T.}},
\bauthor{\bsnm{Wang}, \binits{J.}}:
\bctitle{A parallel derivation of probabilistic information retrieval models}.
In: \bbtitle{Proceedings of the 29th Annual International ACM SIGIR Conference
  on Research and Development in Information Retrieval}.
\bsertitle{SIGIR ’06},
pp. \bfpage{107}--\blpage{114}.
\bpublisher{Association for Computing Machinery},
\blocation{New York, NY, USA}
(\byear{2006}).
\doiurl{10.1145/1148170.1148192}
\end{bchapter}
\endbibitem

\bibitem{Roelleke2008}
\begin{bchapter}
\bauthor{\bsnm{Roelleke}, \binits{T.}},
\bauthor{\bsnm{Wang}, \binits{J.}}:
\bctitle{{TF-IDF} uncovered: A study of theories and probabilities}.
In: \bbtitle{Proceedings of the 31st Annual International ACM SIGIR Conference
  on Research and Development in Information Retrieval}.
\bsertitle{SIGIR '08},
pp. \bfpage{435}--\blpage{442}.
\bpublisher{ACM},
\blocation{New York, NY, USA}
(\byear{2008}).
\doiurl{10.1145/1390334.1390409}
\end{bchapter}
\endbibitem

\bibitem{Roelleke2013}
\begin{bbook}
\bauthor{\bsnm{Roelleke}, \binits{T.}}:
\bbtitle{Information Retrieval Models: Foundations and Relationships}.
\bpublisher{Morgan \& Claypool Publishers},
\blocation{San Rafael, Calif. (1537 Fourth Street, San Rafael, CA 94901 USA)}
(\byear{2013}).
\doiurl{10.1007/978-3-031-02328-6}
\end{bbook}
\endbibitem

\bibitem{Havrlant2017}
\begin{barticle}
\bauthor{\bsnm{Havrlant}, \binits{L.}},
\bauthor{\bsnm{Kreinovich}, \binits{V.}}:
\batitle{A simple probabilistic explanation of term frequency-inverse document
  frequency (tf-idf) heuristic (and variations motivated by this explanation)}.
\bjtitle{International Journal of General Systems}
\bvolume{46}(\bissue{1}),
\bfpage{27}--\blpage{36}
(\byear{2017}).
\doiurl{10.1080/03081079.2017.1291635}
\end{barticle}
\endbibitem

\bibitem{Rivals2007}
\begin{barticle}
\bauthor{\bsnm{Rivals}, \binits{I.}},
\bauthor{\bsnm{Personnaz}, \binits{L.}},
\bauthor{\bsnm{Taing}, \binits{L.}},
\bauthor{\bsnm{Potier}, \binits{M.-C.}}:
\batitle{Enrichment or depletion of a {GO} category within a class of genes:
  Which test?}
\bjtitle{Bioinformatics}
\bvolume{23}(\bissue{4}),
\bfpage{401}--\blpage{407}
(\byear{2007}).
\doiurl{10.1093/bioinformatics/btl633}
\end{barticle}
\endbibitem

\bibitem{Boyle2004}
\begin{barticle}
\bauthor{\bsnm{Boyle}, \binits{E.I.}},
\bauthor{\bsnm{Weng}, \binits{S.}},
\bauthor{\bsnm{Gollub}, \binits{J.}},
\bauthor{\bsnm{Jin}, \binits{H.}},
\bauthor{\bsnm{Botstein}, \binits{D.}},
\bauthor{\bsnm{Cherry}, \binits{J.M.}},
\bauthor{\bsnm{Sherlock}, \binits{G.}}:
\batitle{{GO::TermFinder}-open source software for accessing {Gene Ontology}
  information and finding significantly enriched {Gene Ontology} terms
  associated with a list of genes}.
\bjtitle{Bioinformatics}
\bvolume{20}(\bissue{18}),
\bfpage{3710}
(\byear{2004}).
\doiurl{10.1093/bioinformatics/bth456}
\end{barticle}
\endbibitem

\bibitem{Huang2009}
\begin{barticle}
\bauthor{\bsnm{Huang}, \binits{D.W.}},
\bauthor{\bsnm{Sherman}, \binits{B.T.}},
\bauthor{\bsnm{Lempicki}, \binits{R.A.}}:
\batitle{Bioinformatics enrichment tools: paths toward the comprehensive
  functional analysis of large gene lists}.
\bjtitle{Nucleic Acids Research}
\bvolume{37}(\bissue{1}),
\bfpage{1}
(\byear{2009}).
\doiurl{10.1093/nar/gkn923}
\end{barticle}
\endbibitem

\bibitem{Maere2005}
\begin{barticle}
\bauthor{\bsnm{Maere}, \binits{S.}},
\bauthor{\bsnm{Heymans}, \binits{K.}},
\bauthor{\bsnm{Kuiper}, \binits{M.}}:
\batitle{{BiNGO}: a {Cytoscape} plugin to assess overrepresentation of {Gene
  Ontology} categories in biological networks}.
\bjtitle{Bioinformatics}
\bvolume{21}(\bissue{16}),
\bfpage{3448}--\blpage{3449}
(\byear{2005}).
\doiurl{10.1093/bioinformatics/bti551}
\end{barticle}
\endbibitem

\bibitem{WardeFarley2010}
\begin{barticle}
\bauthor{\bsnm{Warde-Farley}, \binits{D.}},
\bauthor{\bsnm{Donaldson}, \binits{S.L.}},
\bauthor{\bsnm{Comes}, \binits{O.}},
\bauthor{\bsnm{Zuberi}, \binits{K.}},
\bauthor{\bsnm{Badrawi}, \binits{R.}},
\bauthor{\bsnm{Chao}, \binits{P.}},
\bauthor{\bsnm{Franz}, \binits{M.}},
\bauthor{\bsnm{Grouios}, \binits{C.}},
\bauthor{\bsnm{Kazi}, \binits{F.}},
\bauthor{\bsnm{Lopes}, \binits{C.T.}},
\bauthor{\bsnm{Maitland}, \binits{A.}},
\bauthor{\bsnm{Mostafavi}, \binits{S.}},
\bauthor{\bsnm{Montojo}, \binits{J.}},
\bauthor{\bsnm{Shao}, \binits{Q.}},
\bauthor{\bsnm{Wright}, \binits{G.}},
\bauthor{\bsnm{Bader}, \binits{G.D.}},
\bauthor{\bsnm{Morris}, \binits{Q.}}:
\batitle{The {GeneMANIA} prediction server: biological network integration for
  gene prioritization and predicting gene function}.
\bjtitle{Nucleic Acids Research}
\bvolume{38}(\bissue{suppl\_2}),
\bfpage{214}--\blpage{220}
(\byear{2010}).
\doiurl{10.1093/nar/gkq537}
\end{barticle}
\endbibitem

\bibitem{Zheng2008}
\begin{barticle}
\bauthor{\bsnm{Zheng}, \binits{Q.}},
\bauthor{\bsnm{Wang}, \binits{X.-J.}}:
\batitle{{GOEAST}: a web-based software toolkit for {Gene Ontology} enrichment
  analysis}.
\bjtitle{Nucleic Acids Research}
\bvolume{36}(\bissue{suppl\_2}),
\bfpage{358}--\blpage{363}
(\byear{2008}).
\doiurl{10.1093/nar/gkn276}
\end{barticle}
\endbibitem

\bibitem{Dermouche2014}
\begin{bchapter}
\bauthor{\bsnm{Dermouche}, \binits{M.}},
\bauthor{\bsnm{Velcin}, \binits{J.}},
\bauthor{\bsnm{Khouas}, \binits{L.}},
\bauthor{\bsnm{Loudcher}, \binits{S.}}:
\bctitle{A joint model for topic-sentiment evolution over time}.
In: \bbtitle{Proceedings of the 2014 IEEE International Conference on Data
  Mining}.
\bsertitle{ICDM '14},
pp. \bfpage{773}--\blpage{778}.
\bpublisher{IEEE Computer Society},
\blocation{Washington, DC, USA}
(\byear{2014}).
\doiurl{10.1109/ICDM.2014.82}
\end{bchapter}
\endbibitem

\bibitem{GlasgowIRG2023}
\begin{botherref}
{G}lasgow {I}nformation {R}etrieval {G}roup: {C}ranfield collection.
\url{http://ir.dcs.gla.ac.uk/resources/test_collections/cran/}
Accessed 2023-05-23
\end{botherref}
\endbibitem

\bibitem{Lang1995}
\begin{bchapter}
\bauthor{\bsnm{Lang}, \binits{K.}}:
\bctitle{News{W}eeder: learning to filter netnews}.
In: \bbtitle{Proceedings of the 12th International Conference on Machine
  Learning},
pp. \bfpage{331}--\blpage{339}.
\bpublisher{Morgan Kaufmann Publishers Inc.},
\blocation{San Francisco, CA, USA}
(\byear{1995})
\end{bchapter}
\endbibitem

\bibitem{Robertson1994}
\begin{bchapter}
\bauthor{\bsnm{Robertson}, \binits{S.E.}},
\bauthor{\bsnm{Walker}, \binits{S.}},
\bauthor{\bsnm{Jones}, \binits{S.}},
\bauthor{\bsnm{Hancock-Beaulieu}, \binits{M.}},
\bauthor{\bsnm{Gatford}, \binits{M.}}:
\bctitle{{O}kapi at {TREC}-3.}
In: \beditor{\bsnm{Harman}, \binits{D.K.}} (ed.)
\bbtitle{TREC}.
\bsertitle{NIST Special Publication},
vol. \bseriesno{500-225},
pp. \bfpage{109}--\blpage{126}.
\bpublisher{National Institute of Standards and Technology (NIST)},
\blocation{Gaithersburg, MD.}
(\byear{1994})
\end{bchapter}
\endbibitem

\bibitem{Robertson2009}
\begin{barticle}
\bauthor{\bsnm{Robertson}, \binits{S.}},
\bauthor{\bsnm{Zaragoza}, \binits{H.}}:
\batitle{The probabilistic relevance framework: {BM}25 and beyond}.
\bjtitle{Found. Trends Inf. Retr.}
\bvolume{3}(\bissue{4}),
\bfpage{333}--\blpage{389}
(\byear{2009}).
\doiurl{10.1561/1500000019}
\end{barticle}
\endbibitem

\bibitem{Lv2011}
\begin{bchapter}
\bauthor{\bsnm{Lv}, \binits{Y.}},
\bauthor{\bsnm{Zhai}, \binits{C.}}:
\bctitle{Lower-bounding term frequency normalization}.
In: \bbtitle{Proceedings of the 20th ACM International Conference on
  Information and Knowledge Management}.
\bsertitle{CIKM '11},
pp. \bfpage{7}--\blpage{16}.
\bpublisher{Association for Computing Machinery},
\blocation{New York, NY, USA}
(\byear{2011}).
\doiurl{10.1145/2063576.2063584}
\end{bchapter}
\endbibitem

\bibitem{Jiminez2018}
\begin{barticle}
\bauthor{\bsnm{Jimenez}, \binits{S.}},
\bauthor{\bsnm{Cucerzan}, \binits{S.-P.}},
\bauthor{\bsnm{Gonzalez}, \binits{F.A.}},
\bauthor{\bsnm{Gelbukh}, \binits{A.}},
\bauthor{\bsnm{Due\~{n}as}, \binits{G.}},
\bauthor{\bsnm{Pinto}, \binits{D.}},
\bauthor{\bsnm{Singh}, \binits{V.K.}},
\bauthor{\bsnm{Villavicencio}, \binits{A.}},
\bauthor{\bsnm{Mayr-Schlegel}, \binits{P.}},
\bauthor{\bsnm{Stamatatos}, \binits{E.}}:
\batitle{{BM}25-{CTF}: Improving {TF} and {IDF} factors in {BM}25 by using
  collection term frequencies}.
\bjtitle{J. Intell. Fuzzy Syst.}
\bvolume{34}(\bissue{5}),
\bfpage{2887}--\blpage{2899}
(\byear{2018}).
\doiurl{10.3233/JIFS-169475}
\end{barticle}
\endbibitem

\bibitem{Joachims1997}
\begin{bchapter}
\bauthor{\bsnm{Joachims}, \binits{T.}}:
\bctitle{A probabilistic analysis of the {Rocchio} algorithm with {TFIDF} for
  text categorization}.
In: \bbtitle{Proceedings of the Fourteenth International Conference on Machine
  Learning}.
\bsertitle{ICML '97},
pp. \bfpage{143}--\blpage{151}.
\bpublisher{Morgan Kaufmann Publishers Inc.},
\blocation{San Francisco, CA, USA}
(\byear{1997})
\end{bchapter}
\endbibitem

\bibitem{Sabbah2017}
\begin{barticle}
\bauthor{\bsnm{Sabbah}, \binits{T.}},
\bauthor{\bsnm{Selamat}, \binits{A.}},
\bauthor{\bsnm{Selamat}, \binits{M.H.}},
\bauthor{\bsnm{Al-Anzi}, \binits{F.S.}},
\bauthor{\bsnm{Herrera-Viedma}, \binits{E.E.}},
\bauthor{\bsnm{Krejcar}, \binits{O.}},
\bauthor{\bsnm{Fujita}, \binits{H.}}:
\batitle{Modified frequency-based term weighting schemes for text
  classification}.
\bjtitle{Appl. Soft Comput.}
\bvolume{58},
\bfpage{193}--\blpage{206}
(\byear{2017}).
\doiurl{10.1016/J.ASOC.2017.04.069}
\end{barticle}
\endbibitem

\bibitem{Kim2019}
\begin{barticle}
\bauthor{\bsnm{Kim}, \binits{S.-W.}},
\bauthor{\bsnm{Gil}, \binits{J.-M.}}:
\batitle{Research paper classification systems based on {TF-IDF} and {LDA}
  schemes}.
\bjtitle{Human-centric Computing and Information Sciences}
\bvolume{9}(\bissue{1}),
\bfpage{30}
(\byear{2019}).
\doiurl{10.1186/s13673-019-0192-7}
\end{barticle}
\endbibitem

\bibitem{Jiang2021}
\begin{barticle}
\bauthor{\bsnm{Jiang}, \binits{Z.}},
\bauthor{\bsnm{Gao}, \binits{B.}},
\bauthor{\bsnm{He}, \binits{Y.}},
\bauthor{\bsnm{Han}, \binits{Y.}},
\bauthor{\bsnm{Doyle}, \binits{P.}},
\bauthor{\bsnm{Zhu}, \binits{Q.}}:
\batitle{Text classification using novel term weighting scheme-based improved
  {TF-IDF} for internet media reports}.
\bjtitle{Mathematical Problems in Engineering}
\bvolume{2021},
\bfpage{1}--\blpage{30}
(\byear{2021}).
\doiurl{10.1155/2021/6619088}
\end{barticle}
\endbibitem

\bibitem{Chawla2023}
\begin{barticle}
\bauthor{\bsnm{Chawla}, \binits{S.}},
\bauthor{\bsnm{Kaur}, \binits{R.}},
\bauthor{\bsnm{Aggarwal}, \binits{P.}}:
\batitle{Text classification framework for short text based on
  {TFIDF}-{F}ast{T}ext}.
\bjtitle{Multimedia Tools and Applications}
(\byear{2023}).
\doiurl{10.1007/s11042-023-15211-5}
\end{barticle}
\endbibitem

\bibitem{Bafna2016}
\begin{bchapter}
\bauthor{\bsnm{{Bafna}}, \binits{P.}},
\bauthor{\bsnm{{Pramod}}, \binits{D.}},
\bauthor{\bsnm{{Vaidya}}, \binits{A.}}:
\bctitle{Document clustering: {TF-IDF} approach}.
In: \bbtitle{2016 International Conference on Electrical, Electronics, and
  Optimization Techniques (ICEEOT)},
pp. \bfpage{61}--\blpage{66}
(\byear{2016}).
\doiurl{10.1109/ICEEOT.2016.7754750}
\end{bchapter}
\endbibitem

\bibitem{Marcinczuk2021}
\begin{bchapter}
\bauthor{\bsnm{Marci{\'n}czuk}, \binits{M.}},
\bauthor{\bsnm{Gniewkowski}, \binits{M.}},
\bauthor{\bsnm{Walkowiak}, \binits{T.}},
\bauthor{\bsnm{Bedkowski}, \binits{M.}}:
\bctitle{Text document clustering: {W}ordnet vs. {TF}-{IDF} vs. word
  embeddings}.
In: \bbtitle{Proceedings of the 11th Global Wordnet Conference},
pp. \bfpage{207}--\blpage{214}.
\bpublisher{Global Wordnet Association},
\blocation{University of South Africa (UNISA)}
(\byear{2021})
\end{bchapter}
\endbibitem

\bibitem{Thielmann2023}
\begin{bchapter}
\bauthor{\bsnm{Thielmann}, \binits{A.}},
\bauthor{\bsnm{Weisser}, \binits{C.}},
\bauthor{\bsnm{Kneib}, \binits{T.}},
\bauthor{\bsnm{Säfken}, \binits{B.}}:
\bctitle{Coherence based document clustering}.
In: \bbtitle{2023 IEEE 17th International Conference on Semantic Computing
  (ICSC)},
pp. \bfpage{9}--\blpage{16}
(\byear{2023}).
\doiurl{10.1109/ICSC56153.2023.00009}
\end{bchapter}
\endbibitem

\bibitem{Firoozeh2020}
\begin{barticle}
\bauthor{\bsnm{Firoozeh}, \binits{N.}},
\bauthor{\bsnm{Nazarenko}, \binits{A.}},
\bauthor{\bsnm{Alizon}, \binits{F.}},
\bauthor{\bsnm{Daille}, \binits{B.}}:
\batitle{Keyword extraction: Issues and methods}.
\bjtitle{Natural Language Engineering}
\bvolume{26}(\bissue{3}),
\bfpage{259}--\blpage{291}
(\byear{2020}).
\doiurl{10.1017/S1351324919000457}
\end{barticle}
\endbibitem

\bibitem{Koloski2021}
\begin{bchapter}
\bauthor{\bsnm{Koloski}, \binits{B.}},
\bauthor{\bsnm{Pollak}, \binits{S.}},
\bauthor{\bsnm{{\v{S}}krlj}, \binits{B.}},
\bauthor{\bsnm{Martinc}, \binits{M.}}:
\bctitle{Extending neural keyword extraction with {TF}-{IDF} tagset matching}.
In: \bbtitle{Proceedings of the EACL Hackashop on News Media Content Analysis
  and Automated Report Generation},
pp. \bfpage{22}--\blpage{29}.
\bpublisher{Association for Computational Linguistics},
\blocation{Online}
(\byear{2021})
\end{bchapter}
\endbibitem

\bibitem{Qian2021}
\begin{barticle}
\bauthor{\bsnm{Qian}, \binits{Y.}},
\bauthor{\bsnm{Jia}, \binits{C.}},
\bauthor{\bsnm{Liu}, \binits{Y.}}:
\batitle{Bert-based text keyword extraction}.
\bjtitle{Journal of Physics: Conference Series}
\bvolume{1992}(\bissue{4}),
\bfpage{042077}
(\byear{2021}).
\doiurl{10.1088/1742-6596/1992/4/042077}
\end{barticle}
\endbibitem

\bibitem{Magdy2020}
\begin{barticle}
\bauthor{\bsnm{Magdy}, \binits{S.}},
\bauthor{\bsnm{Abouelseoud}, \binits{Y.}},
\bauthor{\bsnm{Mikhail}, \binits{M.}}:
\batitle{Privacy preserving search index for image databases based on {SURF}
  and order preserving encryption}.
\bjtitle{IET Image Processing}
\bvolume{14}(\bissue{5}),
\bfpage{874}--\blpage{881}
(\byear{2020}).
\doiurl{10.1049/iet-ipr.2019.0575}
\end{barticle}
\endbibitem

\bibitem{Yang2007}
\begin{bchapter}
\bauthor{\bsnm{Yang}, \binits{J.}},
\bauthor{\bsnm{Jiang}, \binits{Y.-G.}},
\bauthor{\bsnm{Hauptmann}, \binits{A.G.}},
\bauthor{\bsnm{Ngo}, \binits{C.-W.}}:
\bctitle{Evaluating bag-of-visual-words representations in scene
  classification}.
In: \bbtitle{Proceedings of the International Workshop on Workshop on
  Multimedia Information Retrieval}.
\bsertitle{MIR '07},
pp. \bfpage{197}--\blpage{206}.
\bpublisher{Association for Computing Machinery},
\blocation{New York, NY, USA}
(\byear{2007}).
\doiurl{10.1145/1290082.1290111}
\end{bchapter}
\endbibitem

\bibitem{Moulin2010}
\begin{bchapter}
\bauthor{\bsnm{Moulin}, \binits{C.}},
\bauthor{\bsnm{Barat}, \binits{C.}},
\bauthor{\bsnm{Ducottet}, \binits{C.}}:
\bctitle{Fusion of tf.idf weighted bag of visual features for image
  classification}.
In: \bbtitle{2010 International Workshop on Content Based Multimedia Indexing
  (CBMI)},
pp. \bfpage{1}--\blpage{6}
(\byear{2010}).
\doiurl{10.1109/CBMI.2010.5529901}
\end{bchapter}
\endbibitem

\bibitem{Suzuki2008}
\begin{bchapter}
\bauthor{\bsnm{Suzuki}, \binits{Y.}},
\bauthor{\bsnm{Mitsukawa}, \binits{M.}},
\bauthor{\bsnm{Kawagoe}, \binits{K.}}:
\bctitle{A image retrieval method using {TFIDF} based weighting scheme}.
In: \bbtitle{2008 19th International Workshop on Database and Expert Systems
  Applications},
pp. \bfpage{112}--\blpage{116}
(\byear{2008}).
\doiurl{10.1109/DEXA.2008.106}
\end{bchapter}
\endbibitem

\bibitem{Kondylidis2018}
\begin{barticle}
\bauthor{\bsnm{Kondylidis}, \binits{N.}},
\bauthor{\bsnm{Tzelepi}, \binits{M.}},
\bauthor{\bsnm{Tefas}, \binits{A.}}:
\batitle{Exploiting {Tf}-{Idf} in deep convolutional neural networks for
  content based image retrieval}.
\bjtitle{Multimedia Tools Appl.}
\bvolume{77}(\bissue{23}),
\bfpage{30729}--\blpage{30748}
(\byear{2018})
\end{barticle}
\endbibitem

\bibitem{Chum2008}
\begin{bchapter}
\bauthor{\bsnm{Chum}, \binits{O.}},
\bauthor{\bsnm{Philbin}, \binits{J.}},
\bauthor{\bsnm{Zisserman}, \binits{A.}}:
\bctitle{Near duplicate image detection: min-hash and tf-idf weighting}.
In: \beditor{\bsnm{Everingham}, \binits{M.}},
\beditor{\bsnm{Needham}, \binits{C.J.}},
\beditor{\bsnm{Fraile}, \binits{R.}} (eds.)
\bbtitle{Proceedings of the British Machine Vision Conference 2008},
pp. \bfpage{1}--\blpage{10}.
\bpublisher{British Machine Vision Association},
\blocation{Leeds, UK}
(\byear{2008}).
\doiurl{10.5244/C.22.50}
\end{bchapter}
\endbibitem

\bibitem{Kaur2022}
\begin{barticle}
\bauthor{\bsnm{Kaur}, \binits{G.}},
\bauthor{\bsnm{Singh}, \binits{N.}},
\bauthor{\bsnm{Kumar}, \binits{M.}}:
\batitle{Image forgery techniques: A review}.
\bjtitle{Artif. Intell. Rev.}
\bvolume{56}(\bissue{2}),
\bfpage{1577}--\blpage{1625}
(\byear{2022}).
\doiurl{10.1007/s10462-022-10211-7}
\end{barticle}
\endbibitem

\bibitem{Shan2019}
\begin{barticle}
\bauthor{\bsnm{Shan}, \binits{W.}},
\bauthor{\bsnm{Yi}, \binits{Y.}},
\bauthor{\bsnm{Huang}, \binits{R.}},
\bauthor{\bsnm{Xie}, \binits{Y.}}:
\batitle{Robust contrast enhancement forensics based on convolutional neural
  networks}.
\bjtitle{Signal Processing: Image Communication}
\bvolume{71},
\bfpage{138}--\blpage{146}
(\byear{2019}).
\doiurl{10.1016/j.image.2018.11.011}
\end{barticle}
\endbibitem

\bibitem{Koul2022}
\begin{barticle}
\bauthor{\bsnm{Koul}, \binits{S.}},
\bauthor{\bsnm{Kumar}, \binits{M.}},
\bauthor{\bsnm{Khurana}, \binits{S.S.}},
\bauthor{\bsnm{Mushtaq}, \binits{F.}},
\bauthor{\bsnm{Kumar}, \binits{K.}}:
\batitle{An efficient approach for copy-move image forgery detection using
  convolution neural network}.
\bjtitle{Multimedia Tools Appl.}
\bvolume{81}(\bissue{8}),
\bfpage{11259}--\blpage{11277}
(\byear{2022}).
\doiurl{10.1007/s11042-022-11974-5}
\end{barticle}
\endbibitem

\bibitem{Walia2022}
\begin{barticle}
\bauthor{\bsnm{Walia}, \binits{S.}},
\bauthor{\bsnm{Kumar}, \binits{K.}},
\bauthor{\bsnm{Kumar}, \binits{M.}}:
\batitle{Unveiling digital image forgeries using markov based quaternions in
  frequency domain and fusion of machine learning algorithms}.
\bjtitle{Multimedia Tools Appl.}
\bvolume{82}(\bissue{3}),
\bfpage{4517}--\blpage{4532}
(\byear{2022}).
\doiurl{10.1007/s11042-022-13610-8}
\end{barticle}
\endbibitem

\bibitem{Bansal2021a}
\begin{bchapter}
\bauthor{\bsnm{Bansal}, \binits{M.}},
\bauthor{\bsnm{Kumar}, \binits{M.}},
\bauthor{\bsnm{Kumar}, \binits{M.}}:
\bctitle{Performance comparison of various feature extraction methods for
  object recognition on {C}altech-101 {I}mage dataset}.
In: \beditor{\bsnm{Choudhary}, \binits{A.}},
\beditor{\bsnm{Agrawal}, \binits{A.P.}},
\beditor{\bsnm{Logeswaran}, \binits{R.}},
\beditor{\bsnm{Unhelkar}, \binits{B.}} (eds.)
\bbtitle{Applications of Artificial Intelligence and Machine Learning},
pp. \bfpage{289}--\blpage{303}.
\bpublisher{Springer},
\blocation{Singapore}
(\byear{2021})
\end{bchapter}
\endbibitem

\bibitem{Bansal2021b}
\begin{barticle}
\bauthor{\bsnm{Bansal}, \binits{M.}},
\bauthor{\bsnm{Kumar}, \binits{M.}},
\bauthor{\bsnm{Sachdeva}, \binits{M.}},
\bauthor{\bsnm{Mittal}, \binits{A.}}:
\batitle{Transfer learning for image classification using {VGG}19: Caltech-101
  image data set}.
\bjtitle{Journal of Ambient Intelligence and Humanized Computing}
\bvolume{14},
\bfpage{3609}--\blpage{3620}
(\byear{2021})
\end{barticle}
\endbibitem

\bibitem{Shaheed2022}
\begin{botherref}
\oauthor{\bsnm{Shaheed}, \binits{K.}},
\oauthor{\bsnm{Mao}, \binits{A.}},
\oauthor{\bsnm{Qureshi}, \binits{I.}},
\oauthor{\bsnm{Kumar}, \binits{M.}},
\oauthor{\bsnm{Hussain}, \binits{S.}},
\oauthor{\bsnm{Ullah}, \binits{I.}},
\oauthor{\bsnm{Zhang}, \binits{X.}}:
{DS}-{CNN}: A pre-trained xception model based on depth-wise separable
  convolutional neural network for finger vein recognition.
Expert Syst. Appl.
\textbf{191}(C)
(2022).
\doiurl{10.1016/j.eswa.2021.116288}
\end{botherref}
\endbibitem

\bibitem{Arensia2012}
\begin{bchapter}
\bauthor{\bsnm{Arnesia}, \binits{P.D.}},
\bauthor{\bsnm{Madenda}, \binits{S.}}:
\bctitle{Matching images with textual document using {TFIDF} method}.
In: \bbtitle{2012 5th International Congress on Image and Signal Processing},
pp. \bfpage{1283}--\blpage{1289}
(\byear{2012}).
\doiurl{10.1109/CISP.2012.6469720}
\end{bchapter}
\endbibitem

\bibitem{Schneider2022}
\begin{bchapter}
\bauthor{\bsnm{Schneider}, \binits{F.}},
\bauthor{\bsnm{Biemann}, \binits{C.}}:
\bctitle{Golden retriever: A real-time multi-modal text-image retrieval system
  with the ability to focus}.
In: \bbtitle{Proceedings of the 45th International ACM SIGIR Conference on
  Research and Development in Information Retrieval}.
\bsertitle{SIGIR '22},
pp. \bfpage{3245}--\blpage{3250}.
\bpublisher{Association for Computing Machinery},
\blocation{New York, NY, USA}
(\byear{2022}).
\doiurl{10.1145/3477495.3531666}
\end{bchapter}
\endbibitem

\bibitem{Xie2022}
\begin{barticle}
\bauthor{\bsnm{Xie}, \binits{Z.}},
\bauthor{\bsnm{Liu}, \binits{L.}},
\bauthor{\bsnm{Wu}, \binits{Y.}},
\bauthor{\bsnm{Li}, \binits{L.}},
\bauthor{\bsnm{Zhong}, \binits{L.}}:
\batitle{Learning {TFIDF} enhanced joint embedding for recipe-image cross-modal
  retrieval service}.
\bjtitle{IEEE Transactions on Services Computing}
\bvolume{15}(\bissue{6}),
\bfpage{3304}--\blpage{3316}
(\byear{2022}).
\doiurl{10.1109/TSC.2021.3098834}
\end{barticle}
\endbibitem

\bibitem{Pavlopoulos2019}
\begin{bchapter}
\bauthor{\bsnm{Pavlopoulos}, \binits{J.}},
\bauthor{\bsnm{Kougia}, \binits{V.}},
\bauthor{\bsnm{Androutsopoulos}, \binits{I.}}:
\bctitle{A survey on biomedical image captioning}.
In: \bbtitle{Proceedings of the Second Workshop on Shortcomings in Vision and
  Language},
pp. \bfpage{26}--\blpage{36}.
\bpublisher{Association for Computational Linguistics},
\blocation{Minneapolis, Minnesota}
(\byear{2019}).
\doiurl{10.18653/v1/W19-1803}
\end{bchapter}
\endbibitem

\bibitem{Ashwath2021}
\begin{bchapter}
\bauthor{\bsnm{Krishnan}, \binits{A.}},
\bauthor{\bsnm{Rajesh}, \binits{S.}},
\bauthor{\bsnm{SS}, \binits{S.}}:
\bctitle{Text-based image retrieval using captioning}.
In: \bbtitle{2021 Fourth International Conference on Electrical, Computer and
  Communication Technologies (ICECCT)},
pp. \bfpage{1}--\blpage{5}
(\byear{2021}).
\doiurl{10.1109/ICECCT52121.2021.9616897}
\end{bchapter}
\endbibitem

\bibitem{Masciari2020}
\begin{bchapter}
\bauthor{\bsnm{Masciari}, \binits{E.}},
\bauthor{\bsnm{Moscato}, \binits{V.}},
\bauthor{\bsnm{Picariello}, \binits{A.}},
\bauthor{\bsnm{Sperl\'{\i}}, \binits{G.}}:
\bctitle{Detecting fake news by image analysis}.
In: \bbtitle{Proceedings of the 24th Symposium on International Database
  Engineering \& Applications}.
\bsertitle{IDEAS '20}.
\bpublisher{Association for Computing Machinery},
\blocation{New York, NY, USA}
(\byear{2020}).
\doiurl{10.1145/3410566.3410599}
\end{bchapter}
\endbibitem

\bibitem{Choras2021}
\begin{barticle}
\bauthor{\bsnm{Choraś}, \binits{M.}},
\bauthor{\bsnm{Demestichas}, \binits{K.}},
\bauthor{\bsnm{Giełczyk}, \binits{A.}},
\bauthor{\bparticle{Álvaro} \bsnm{Herrero}},
\bauthor{\bsnm{Ksieniewicz}, \binits{P.}},
\bauthor{\bsnm{Remoundou}, \binits{K.}},
\bauthor{\bsnm{Urda}, \binits{D.}},
\bauthor{\bsnm{Woźniak}, \binits{M.}}:
\batitle{Advanced machine learning techniques for fake news (online
  disinformation) detection: A systematic mapping study}.
\bjtitle{Applied Soft Computing}
\bvolume{101},
\bfpage{107050}
(\byear{2021}).
\doiurl{10.1016/j.asoc.2020.107050}
\end{barticle}
\endbibitem

\bibitem{Mangolin2022}
\begin{barticle}
\bauthor{\bsnm{Mangolin}, \binits{R.B.}},
\bauthor{\bsnm{Pereira}, \binits{R.M.}},
\bauthor{\bsnm{Britto}, \binits{A.S.}},
\bauthor{\bsnm{Silla}, \binits{C.N.}},
\bauthor{\bsnm{Feltrim}, \binits{V.D.}},
\bauthor{\bsnm{Bertolini}, \binits{D.}},
\bauthor{\bsnm{Costa}, \binits{Y.M.G.}}:
\batitle{A multimodal approach for multi-label movie genre classification}.
\bjtitle{Multimedia Tools and Applications}
\bvolume{81}(\bissue{14}),
\bfpage{19071}--\blpage{19096}
(\byear{2022}).
\doiurl{10.1007/s11042-020-10086-2}
\end{barticle}
\endbibitem

\bibitem{Rajput2022}
\begin{barticle}
\bauthor{\bsnm{Rajput}, \binits{N.K.}},
\bauthor{\bsnm{Grover}, \binits{B.A.}}:
\batitle{A multi-label movie genre classification scheme based on the movie's
  subtitles}.
\bjtitle{Multimedia Tools and Applications}
\bvolume{81}(\bissue{22}),
\bfpage{32469}--\blpage{32490}
(\byear{2022}).
\doiurl{10.1007/s11042-022-12961-6}
\end{barticle}
\endbibitem

\bibitem{Giveki2021}
\begin{barticle}
\bauthor{\bsnm{Giveki}, \binits{D.}}:
\batitle{Scale-space multi-view bag of words for scene categorization}.
\bjtitle{Multimedia Tools and Applications}
\bvolume{80}(\bissue{1}),
\bfpage{1223}--\blpage{1245}
(\byear{2021}).
\doiurl{10.1007/s11042-020-09759-9}
\end{barticle}
\endbibitem

\bibitem{Kannao2022}
\begin{barticle}
\bauthor{\bsnm{Kannao}, \binits{R.}},
\bauthor{\bsnm{Guha}, \binits{P.}},
\bauthor{\bsnm{Chaudhuri}, \binits{B.B.}}:
\batitle{Only overlay text: novel features for {TV} news broadcast video
  segmentation}.
\bjtitle{Multimedia Tools and Applications}
\bvolume{81}(\bissue{21}),
\bfpage{30493}--\blpage{30517}
(\byear{2022}).
\doiurl{10.1007/s11042-022-12917-w}
\end{barticle}
\endbibitem

\bibitem{Devlin2019}
\begin{bchapter}
\bauthor{\bsnm{Devlin}, \binits{J.}},
\bauthor{\bsnm{Chang}, \binits{M.-W.}},
\bauthor{\bsnm{Lee}, \binits{K.}},
\bauthor{\bsnm{Toutanova}, \binits{K.}}:
\bctitle{{BERT}: Pre-training of deep bidirectional transformers for language
  understanding}.
In: \bbtitle{Proceedings of the 2019 Conference of the North {A}merican Chapter
  of the Association for Computational Linguistics: Human Language
  Technologies, Volume 1 (Long and Short Papers)},
pp. \bfpage{4171}--\blpage{4186}.
\bpublisher{Association for Computational Linguistics},
\blocation{Minneapolis, Minnesota}
(\byear{2019}).
\doiurl{10.18653/v1/N19-1423}
\end{bchapter}
\endbibitem

\bibitem{vonderMosel2022}
\begin{barticle}
\bauthor{\bparticle{von~der} \bsnm{Mosel}, \binits{J.}},
\bauthor{\bsnm{Trautsch}, \binits{A.}},
\bauthor{\bsnm{Herbold}, \binits{S.}}:
\batitle{On the validity of pre-trained transformers for natural language
  processing in the software engineering domain}.
\bjtitle{IEEE Transactions on Software Engineering}
\bvolume{49}(\bissue{4}),
\bfpage{1487}--\blpage{1507}
(\byear{2023}).
\doiurl{10.1109/TSE.2022.3178469}
\end{barticle}
\endbibitem

\bibitem{Robertson1974}
\begin{barticle}
\bauthor{\bsnm{Robertson}, \binits{S.}}:
\batitle{Specificity and weighted retrieval}.
\bjtitle{Journal of Documentation}
\bvolume{30}(\bissue{1}),
\bfpage{41}--\blpage{46}
(\byear{1974})
\end{barticle}
\endbibitem

\bibitem{Wong1992}
\begin{barticle}
\bauthor{\bsnm{Wong}, \binits{S.K.M.}},
\bauthor{\bsnm{Yao}, \binits{Y.}}:
\batitle{An information-theoretic measure of term specificity}.
\bjtitle{Journal of the American Society for Information Science}
\bvolume{43}(\bissue{1}),
\bfpage{54}--\blpage{61}
(\byear{1992}).
\doiurl{10.1002/(SICI)1097-4571(199201)43:1<54::AID-ASI5>3.0.CO;2-A}
\end{barticle}
\endbibitem

\bibitem{Robertson1976}
\begin{barticle}
\bauthor{\bsnm{Robertson}, \binits{S.}},
\bauthor{\bsnm{Jones}, \binits{K.S.}}:
\batitle{Relevance weighting of search terms}.
\bjtitle{Journal of the American Society for Information Science}
\bvolume{27}(\bissue{3}),
\bfpage{129}--\blpage{146}
(\byear{1976}).
\doiurl{10.1002/asi.4630270302}
\end{barticle}
\endbibitem

\bibitem{Manning2008}
\begin{bbook}
\bauthor{\bsnm{Manning}, \binits{C.D.}},
\bauthor{\bsnm{Raghavan}, \binits{P.}},
\bauthor{\bsnm{Sch\"{u}tze}, \binits{H.}}:
\bbtitle{Introduction to Information Retrieval},
pp. \bfpage{227}--\blpage{228}.
\bpublisher{Cambridge University Press},
\blocation{New York, NY, USA}
(\byear{2008}).
\bcomment{Chap. 11}.
\doiurl{10.1017/CBO9780511809071}
\end{bbook}
\endbibitem

\bibitem{Wu2008}
\begin{barticle}
\bauthor{\bsnm{Wu}, \binits{H.C.}},
\bauthor{\bsnm{Luk}, \binits{R.W.P.}},
\bauthor{\bsnm{Wong}, \binits{K.F.}},
\bauthor{\bsnm{Kwok}, \binits{K.L.}}:
\batitle{Interpreting {TF-IDF} term weights as making relevance decisions}.
\bjtitle{ACM Transactions on Information Systems}
\bvolume{26}(\bissue{3}),
\bfpage{13}--\blpage{11337}
(\byear{2008}).
\doiurl{10.1145/1361684.1361686}
\end{barticle}
\endbibitem

\bibitem{Dua2023}
\begin{botherref}
\oauthor{\bsnm{Dua}, \binits{D.}},
\oauthor{\bsnm{Graff}, \binits{C.}}:
{UCI} Machine Learning Repository
(2023).
\url{https://archive-beta.ics.uci.edu}
\end{botherref}
\endbibitem

\bibitem{Oussama2022}
\begin{botherref}
\oauthor{\bsnm{Oussama}, \binits{B.K.}}:
Cranfield collection in {TREC} {XML} format.
GitHub
(2022).
\url{https://github.com/oussbenk/cranfield-trec-dataset}
\end{botherref}
\endbibitem

\bibitem{Pedregosa2011}
\begin{barticle}
\bauthor{\bsnm{Pedregosa}, \binits{F.}},
\bauthor{\bsnm{Varoquaux}, \binits{G.}},
\bauthor{\bsnm{Gramfort}, \binits{A.}},
\bauthor{\bsnm{Michel}, \binits{V.}},
\bauthor{\bsnm{Thirion}, \binits{B.}},
\bauthor{\bsnm{Grisel}, \binits{O.}},
\bauthor{\bsnm{Blondel}, \binits{M.}},
\bauthor{\bsnm{Prettenhofer}, \binits{P.}},
\bauthor{\bsnm{Weiss}, \binits{R.}},
\bauthor{\bsnm{Dubourg}, \binits{V.}},
\bauthor{\bsnm{Vanderplas}, \binits{J.}},
\bauthor{\bsnm{Passos}, \binits{A.}},
\bauthor{\bsnm{Cournapeau}, \binits{D.}},
\bauthor{\bsnm{Brucher}, \binits{M.}},
\bauthor{\bsnm{Perrot}, \binits{M.}},
\bauthor{\bsnm{Duchesnay}, \binits{E.}}:
\batitle{Scikit-learn: Machine learning in {P}ython}.
\bjtitle{Journal of Machine Learning Research}
\bvolume{12},
\bfpage{2825}--\blpage{2830}
(\byear{2011})
\end{barticle}
\endbibitem

\bibitem{Irvine2017}
\begin{barticle}
\bauthor{\bsnm{Irvine}, \binits{A.}},
\bauthor{\bsnm{Callison-Burch}, \binits{C.}}:
\batitle{A comprehensive analysis of bilingual lexicon induction}.
\bjtitle{Computational Linguistics}
\bvolume{43}(\bissue{2}),
\bfpage{273}--\blpage{310}
(\byear{2017}).
\doiurl{10.1162/COLI_a_00284}
\end{barticle}
\endbibitem

\bibitem{Cao2014}
\begin{barticle}
\bauthor{\bsnm{Cao}, \binits{J.}},
\bauthor{\bsnm{Zhang}, \binits{S.}}:
\batitle{A bayesian extension of the hypergeometric test for functional
  enrichment analysis}.
\bjtitle{Biometrics}
\bvolume{70}(\bissue{1}),
\bfpage{84}--\blpage{94}
(\byear{2014}).
\doiurl{10.1111/biom.12122}
\end{barticle}
\endbibitem

\bibitem{Cao2017}
\begin{barticle}
\bauthor{\bsnm{Cao}, \binits{J.}}:
\batitle{Bayesian functional enrichment analysis for the {R}eactome database}.
\bjtitle{Statistical Theory and Related Fields}
\bvolume{1}(\bissue{2}),
\bfpage{185}--\blpage{193}
(\byear{2017}).
\doiurl{10.1080/24754269.2017.1387444}
\end{barticle}
\endbibitem

\bibitem{Fan2021}
\begin{barticle}
\bauthor{\bsnm{Fan}, \binits{R.}},
\bauthor{\bsnm{Cui}, \binits{Q.}}:
\batitle{Toward comprehensive functional analysis of gene lists weighted by
  gene essentiality scores}.
\bjtitle{Bioinformatics}
\bvolume{37}(\bissue{23}),
\bfpage{4399}--\blpage{4404}
(\byear{2021}).
\doiurl{10.1093/bioinformatics/btab475}
\end{barticle}
\endbibitem

\bibitem{Onsjo2020}
\begin{barticle}
\bauthor{\bsnm{Onsj{\"o}}, \binits{M.}},
\bauthor{\bsnm{Sheridan}, \binits{P.}}:
\batitle{Theme enrichment analysis: A statistical test for identifying
  significantly enriched themes in a list of stories with an application to the
  {Star Trek} television franchise}.
\bjtitle{Digital Studies/le champ num{\'e}rique}
\bvolume{10}(\bissue{1}),
\bfpage{1}
(\byear{2020}).
\doiurl{10.16995/dscn.316}
\end{barticle}
\endbibitem

\end{thebibliography}


\end{document}